\documentclass[doublecol,figures]{epl2} 
\usepackage{amsmath}
\usepackage{amssymb}
\DeclareMathOperator*{\argmax}{argmax}

\title{Dynamical transitions in a driven diffusive model with interactions}
\author{D. Botto\inst{1,2}\thanks{E-mail: \email{davide.botto@polito.it}} \and A. Pelizzola\inst{1,2}\thanks{E-mail: \email{alessandro.pelizzola@polito.it}} \and M. Pretti\inst{1,3}\thanks{E-mail: \email{marco.pretti@polito.it}}}
\shortauthor{D. Botto \etal}

\institute{ \inst{1} Dipartimento di Scienza Applicata e Tecnologia,
  Politecnico di Torino, Corso Duca degli Abruzzi 24, I--10129 Torino,
  Italy\\ 
\inst{2} INFN, Sezione di Torino, via Pietro Giuria 1,
  I-10125 Torino, Italy\\ 
\inst{3} Consiglio Nazionale delle Ricerche
  - Istituto dei Sistemi Complessi (CNR-ISC), Via dei Taurini 19,
  00185 Roma, Italy } 

\abstract{
We study the dynamics of an asymmetric simple exclusion process with open boundaries and local interactions using a pair approximation which generalizes the 2--node cluster mean field theory and the Markov chain approach to kinetics and shares with these approaches the property of reproducing exact results for the bulk current--density relation and the steady state phase diagrams. We find that the relaxation rate exhibits a dynamical transition, with no {static} counterpart, analogous to that {found without} interactions. Remarkably, for some values of the model's parameters, we find 2 dynamical transitions in the same low density phase. We study the dynamics of relaxation to the steady state on both sides of these transitions and make an attempt at providing a physical interpretation for this phenomenon. Results from numerical approaches and a modified Domain Wall Theory confirm the picture provided by the pair approximation.
}

\begin{document}

\maketitle

\section{Introduction}

A fundamental aspect of non--equilibrium statistical physics is the investigation of steady states (SS) \cite{vanKampen}, which are not yet as well understood as their equilibrium counterparts. Driven lattice gases have been shown to be excellent model systems for such investigations, and a prominent role in this class of models is played by the Asymmetric Simple Exclusion Process (ASEP) and its generalizations, inspired by biological and vehicular traffic phenomena (see \cite{ZiaReview,TransportBook} for reviews). In ASEP, the nodes of a one--dimensional lattice can be occupied by at most one particle, and particles hop to empty nearest--neighbour nodes with asymmetric rates, e.g.\ hopping in the rightward direction is more likely than leftward hopping. If leftward hopping is forbidden the model is called Totally Asymmetric Simple Exclusion Process (TASEP). On an open lattice, injection and extraction of particles are allowed at lattice boundaries, and the SS of the models exhibits, as a function of the injection and extraction rates, rich phase diagrams, well described by the theory of boundary--induced phase transitions \cite{Schutz}. In the last decades, many exact results have been obtained \cite{DerridaDomanyMukamel92,SchutzDomany93,Derrida-etal93,Derrida98} for these phase diagrams and other properties such as density profiles.  These models have thus become paradigmatic in non--equilibrium statistical physics, like the Ising model in the equilibrium case. 

In an attempt at moving towards more realistic modeling of vehicular traffic, Antal and Sch\"utz (AS) considered a TASEP with local interactions \cite{AS}. In the AS model, rates depend on the occupation of the next--nearest--neighbour node in the direction of motion. The model is an instance of a more general one, previously introduced by Katz, Lebowitz and Spohn (KLS) \cite{KLS}. Both attractive and repulsive interactions were considered, leading to different physical behaviours. Among many results in \cite{AS}, it is worth mentioning an exact solution for the SS distribution in special cases. In particular, for periodic boundary conditions, and also for open boundary conditions with special (bulk--adapted) values of the boundary rates, the SS distribution can be written as the equilibrium distribution of a one--dimensional Ising model with nearest--neighbour interactions, a property shared by several models in the KLS class. Correlations in this SS are therefore richer than those exhibited by 
%certain simple exclusion processes like 
TASEP, whose SS distribution, under appropriate conditions, factors over nodes in a mean--field like fashion, making certain mean--field results (e.g.\ the location of many SS phase transitions) exact. Indeed, AS reported a very poor performance of mean--field for the SS properties of their model. Given that the pair approximation (PA) is exact for the equilibrium one--dimensional Ising model (see e.g.\ \cite{MyRev} and refs.\ therein), various PAs have been recently employed with success \cite{Maass2011,Maass2012,Maass2013,Kolomeisky2015A,Kolomeisky2015B,Baumgaertner2017,Kolomeisky2018A,Kolomeisky2018B} in the study of several models in the KLS class and their generalizations. It is therefore worth investigating how a PA performs in the case of the AS model, and applying it to the study of properties of this model for which an exact solution is not available. 

In this direction, it is of particular interest to consider the possibility of existence of a dynamical transition, which will be the main focus of this work. This transition, found and exactly located by de Gier and Essler \cite{deGierEssler05,deGierEssler08} in certain ASEPs, including TASEP, corresponds to a singularity in the relaxation rate which is not associated to any singularity in the 
SS. In spite of the fact that the location of the transition is exactly known, its physical meaning is not yet well understood \cite{ProemeBlytheEvans11}. In the case of TASEP, the transition has been recently found, and located with reasonable accuracy, in the framework of different mean--field like approximations of increasing complexity, including a PA \cite{TASEP-EPJB}. 

The aim of the present paper is therefore to show, using a PA, supported by results from numerical approaches and a modified Domain Wall Theory (mDWT), that the dynamical transition is a robust phenomenon, which is exhibited also in the case of the AS model, and to make an attempt at providing a physical interpretation. 

\section{Model and pair approximation}

The AS model \cite{AS} is defined on a one--dimensional lattice of $N$ nodes, labeled $i = 1, 2, \ldots , N$, with open boundaries. Each node can be empty or singly occupied, the %corresponding 
occupation number variable for node $i$ at time $t$ is $n_i^t = 0, 1$. In the following we will denote by $P_i^t[n_i n_{i+1} \cdots n_{i+k}]$ the probability that, at time $t$, the occupation numbers of nodes from $i$ to $i+k$ take values $n_i, n_{i+1}, \cdots, n_{i+k}$ respectively. The average of an occupation number variable is the local density $\rho_i^t = \langle n_i^t \rangle = P_i^t[1]$. In the SS local densities do not depend on time and are denoted by $\rho_i$, dropping the time index. If the local density SS is also uniform, we denote it simply by $\rho$, dropping also the node index. A particle at node $i$ can hop to node $i+1$, provided this is empty, with a rate which depends on the occupation of node $i+2$. If node $i+2$ is empty (respectively occupied), the hopping rate from $i$ to $i+1$ is denoted by $r$ (resp.\ $q$). For $q < r$ (respectively $q > r$) interactions are said to be repulsive (resp.\ attractive). The current $J_i^t$ from node $i$ to node $i+1$ at time $t$ can be written as
\begin{eqnarray}
J_i^t &=& \langle n_i^t (1 - n_{i+1}^t) [ q n_{i+2}^t + r(1 - n_{i+2}^t)] \rangle \nonumber \\ &=& q P_i^t[101] + r P_i^t[100], \qquad i = 1, \ldots , N-2.
\label{eq:Jit}
\end{eqnarray}
It was shown in \cite{AS} that a model with the kinetics described above and periodic boundary conditions has a SS current--density relation in the thermodynamical limit given by
\begin{equation}
J(\rho) = r \rho \left[ 1 + \frac{\sqrt{1 - 4 \rho(1 - \rho)(1 - q/r)}-1}{2(1 - \rho)(1 - q/r)} \right].
\label{eq:J}
\end{equation}
On a lattice with open boundaries, some care is needed in the definition of the boundary rates. In \cite{AS} these rates have been defined in such a way that they would yield constant density profiles for semi--infinite systems (these boundary rates are usually called bulk--adapted \cite{Maass2012,Maass2013,Kolomeisky2018A}, whereas a possible different choice is that of equilibrated--bath \cite{Maass2012,Maass2013} boundary rates). Consider the left boundary: it is reasonable to assume that the injection rate at node $1$ depends on the occupation of node $2$. This injection rate is denoted by $\alpha_1$ (respectively $\alpha_2$) if node $2$ is occupied (resp.\ empty). It has been shown in \cite{AS} that imposing the condition that a uniform density $\rho_L$ is obtained in the SS of a semi--infinite system ($i = 1, 2, \ldots, \infty$) one obtains
\begin{equation}
\alpha_1 = q \left[ 1 - \frac{J(\rho_L)}{r \rho_L} \right], \qquad \alpha_2 = r \left[ 1 - \frac{J(\rho_L)}{r \rho_L} \right].
\label{eq:alpha12}
\end{equation}
Consider now the right boundary: here one needs to specify the hopping rate from node $N-1$ to node $N$, which is denoted by $\beta_1$, and the extraction rate from node $N$, denoted by $\beta_2$. The condition that a uniform density $\rho_R$ is obtained in the SS of a semi--infinite system ($i = -\infty, \ldots , N-1, N$) now gives \cite{AS}
\begin{equation}
\beta_1 = \frac{J(\rho_R)}{1-\rho_R} \left[ 1 - \frac{J(\rho_R)}{r \rho_R} \right]^{-1}, \qquad \beta_2 = \frac{J(\rho_R)}{\rho_R}.
\label{eq:beta12}
\end{equation}
%Notice that in the limit $q \to 1$, $r \to 1$ we obtain $\alpha_1 = \alpha_2 = \rho_L$, $\beta_1 = 1$ and $\beta_2 = 1 - \rho_R$, and pure TASEP is recovered, with $\rho_L$ and $1 - \rho_R$ corresponding to the injection and extraction rates respectively. 

In order to introduce the PA we will assume, as in previous works based on the Markov chain approach to kinetics (MCAK) \cite{Maass2011,Maass2012,Maass2013}, the cluster mean--field (CMF) theory \cite{Kolomeisky2018A,Kolomeisky2018B} and related ideas \cite{nm,Ito1995,SchweitzerBehera2015,TASEP-EPJB}, that $k$--node marginals ($k \ge 3$) factor, at any given time $t$, according to
\begin{equation}
P_i^t[n_i n_{i+1} \ldots n_{i+k-1}] = \frac{\prod_{l=i}^{i+k-2} P_l^t[n_l n_{l+1}]}{\prod_{l=i+1}^{i+k-2} P_l^t[n_l]}.
\label{eq:Pair}
\end{equation}

The 2--node marginal $P_i^t[n_i n_{i+1}]$ ($i = 1, \ldots , N-1$), exploiting normalization, can be written in terms of 3 parameters: in the following we will use as parameters the 2 local densities $\rho_i^t$ and $\rho_{i+1}^t$ together with $\psi_i^t = P_i^t[10]$. As a consequence we have $P_i^t[00] = 1 - \rho_{i+1}^t - \psi_i^t$, $P_i^t[01] = \rho_{i+1}^t - \rho_i^t + \psi_i^t$ and $P_i^t[11] = \rho_i^t - \psi_i^t$. 
With the above assumptions, we can now write the equation for the time evolution of $\rho_i^t$ and $\psi_i^t$. For the local densities we obtain
\begin{equation}
\dot \rho_i^t = J_{i-1}^t - J_i^t, \qquad i = 1, \ldots , N,
\label{eq:rhodot}
\end{equation}
where the current in the PA is given by Eq.\ \ref{eq:Jit} with Eq.\ \ref{eq:Pair} 
%\begin{equation}
%P_i^t[101] = \frac{\psi_i^t (\rho_{i+2}^t - \rho_{i+1}^t + \psi_{i+1}^t)}{1 - \rho_{i+1}^t}
%\label{eq:P101}
%\end{equation}
and
%\begin{equation}
%P_i^t[100] = \frac{\psi_i^t (1 - \rho_{i+2}^t - \psi_{i+1}^t)}{1 - \rho_{i+1}^t}.
%\label{eq:P100}
%\end{equation}
boundary currents are given by
\begin{eqnarray}
J_0^t &=& \alpha_1 P_1^t[01] + \alpha_2 P_1^t[00], \nonumber \\
J_{N-1}^t &=& \beta_1 P_{N-1}^t[10], \nonumber \\
J_N^t &=& \beta_2 P_N^t[1]. 
\label{eq:J0NPair}
\end{eqnarray}
For the 2--node expectations we obtain
\begin{equation}
\dot \psi_i^t = r P_{i-1}^t[100] + q P_i^t[1101] + r P_i^t[1100] - J_i^t
\label{eq:pidot}
\end{equation}
for $i = 2, \ldots , N-3$ and
\begin{eqnarray}
\dot \psi_1^t &=& \alpha_2 P_1^t[00] + q P_1^t[1101] + r P_1^t[1100] - J_1^t, \nonumber \\
\dot \psi_{N-2}^t &=& r P_{N-3}^t[100] + \beta_1 P_{N-2}^t[110] - J_{N-2}^t, \nonumber \\
\dot \psi_{N-1}^t &=& r P_{N-2}^t[100] + \beta_2 P_{N-1}^t[11] - J_{N-1}^t
\label{eq:pidot1N}
\end{eqnarray}
at the boundaries. Eqs.\ \ref{eq:pidot}--\ref{eq:pidot1N} represent an improvement with respect to the MCAK \cite{Maass2011,Maass2012,Maass2013}, where the dynamical equations are closed by assuming that the 2--node expectations, or correlators, $\psi_i^t$ depend at any time on the local densities in the same way as they do in the equilibrium one--dimensional Ising model describing the SS.

Notice that Eqs.\ \ref{eq:rhodot}--\ref{eq:pidot1N} can be viewed, by expressing 3-- and 4--node marginals using Eq.\ \ref{eq:Pair}, as an equation 
\begin{equation}
\dot x^t = f(x^t)
\label{eq:xdot}
\end{equation}
for the time evolution of the $(2N-1)$--component vector
\begin{equation}
x^t = (\rho_1^t, \psi_1^t, \ldots , \rho_{N-1}^t, \psi_{N-1}^t, \rho_N^t).
\label{eq:x}
\end{equation}
%of our dynamical variables. 
The SS $x = (\rho_1, \psi_1, \ldots , \rho_{N-1}, \psi_{N-1}, \rho_N)$ will be given by the condition $f(x) = 0$, and relaxation near the SS will be described by the relaxation matrix $M$, with elements %are defined by 
\begin{equation}
M_{ab} = - \left. \frac{\partial f_a}{\partial x_b^t} \right\vert_{x^t = x}, \qquad a, b = 1, \ldots , 2N-1.
\label{eq:Relax}
\end{equation}
In particular, its smallest eigenvalue $\lambda_1$ is the slowest relaxation rate, the inverse of the longest relaxation time.

\section{Results}

First of all, we look for bulk solutions in the SS, where by continuity the current is uniform, $J_i = J$. In more detail, we look for a SS with $\rho_i = \rho$ and $\psi_i = \psi$ (as a consequence all marginals will be independent of position), at least sufficiently far from the boundaries. In this case the condition $\dot \psi_i^t = 0$ becomes (dropping indices $i$ and $t$ in the marginals)
\begin{eqnarray}
0 &=& r P[100] + q P[1101] + r P[1100] \nonumber \\
&& - (q P[101] + r P[100]) \nonumber \\
&=& r P[1100] - q P[0101] \nonumber \\
&=& r \frac{(\rho - \psi) \psi (1 - \rho - \psi)}{\rho (1 - \rho)} - q \frac{\psi^3}{\rho (1 - \rho)},
\label{eq:Bulk}
\end{eqnarray}
which is solved by 
\begin{equation}
\psi = \frac{1 - \sqrt{1 - 4 \rho (1 - \rho) (1 - q/r)}}{2 (1 - q/r)}.
\label{eq:piBulk}
\end{equation}
The corresponding current is
\begin{eqnarray}
J(\rho) &=& q P[101] + r P[100] \nonumber \\
&=& q \frac{\psi^2}{1 - \rho} + r \frac{\psi (1 - \rho - \psi)}{1 - \rho} \nonumber \\
&=& r \rho \left( 1 - \frac{\psi}{1 - \rho} \right),
\label{eq:JBulk}
\end{eqnarray}
which turns out to be exact (see Eq.\ \ref{eq:J} and \cite{AS}). Simple algebra shows that with the boundary rates defined as in Eqs.\ \ref{eq:alpha12}--\ref{eq:beta12} with $\rho_L = \rho_R = \rho$, in the bulk SS Eqs.\ \ref{eq:J0NPair} yield $J_0 = J_{N-1} = J_N = J(\rho)$ and the r.h.s.\ of Eqs.\ \ref{eq:pidot1N} vanish. With this definition of the boundary rates, in the PA we find a bulk SS with the exact current--density relation at any finite size $N$. This implies that in the PA the exact location of most SS phase transitions is recovered. These exact results, and as a consequence the location of most transition lines in the phase diagram, can also be obtained by using the CMF theory in \cite{Kolomeisky2018A,Kolomeisky2018B} or the MCAK \cite{Maass2011,Maass2012,Maass2013}.

\begin{figure}
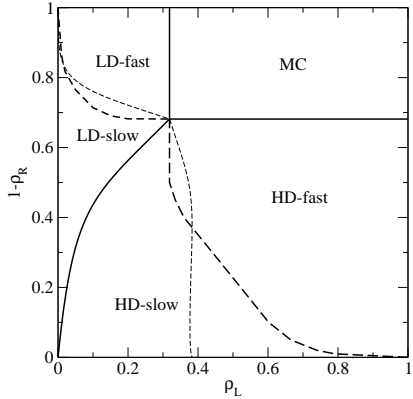

\onefigure[width=0.3\textwidth]{Fig1.eps}
\caption{Typical phase diagram for repulsive interactions, here $q = 0.1$ and $r = 1$. Solid lines denote SS transitions. Thick (respectively thin) dashed lines denoted dynamical transitions given by the PA (resp.\ mDWT). Phase labels are explained in the text. }
\label{fig:PhDiagRep}
\end{figure}

\begin{figure}
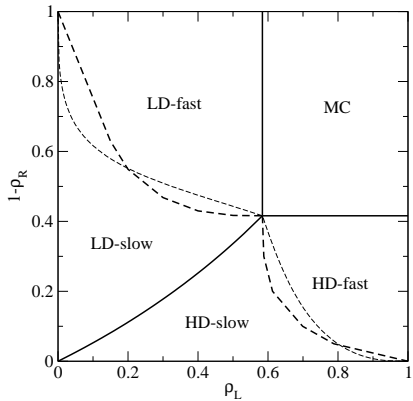

\onefigure[width=0.3\textwidth]{Fig2.eps}
\caption{Same as Fig.\ \protect\ref{fig:PhDiagRep} for weakly attractive interactions, here $q = 1$ and $r = 0.5$.}
\label{fig:PhDiagAttr}
\end{figure}

\begin{figure}
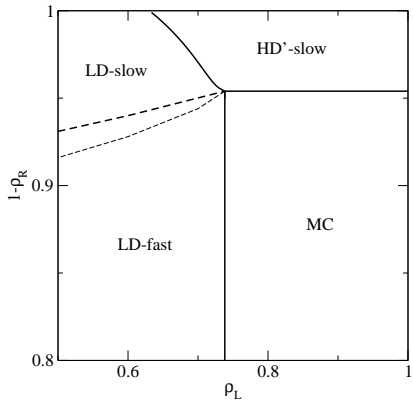

\onefigure[width=0.3\textwidth]{Fig3.eps}
\caption{Same as Fig.\ \protect\ref{fig:PhDiagRep} for strongly attractive interactions, here $q = 1$ and $r = 0.1$. The portion of the phase diagram with the HD$'$ phase is shown.}
\label{fig:PhDiagAttr10}
\end{figure}

Let us focus on the SS phase diagram, in the limit of large lattice size $N$, using as parameters, in addition to $q$ and $r$, the densities $\rho_L$ and $\rho_R$. More precisely, in order to make contact with \cite{AS} and the literature on TASEP, our parameters will be $\rho_L$ and $1 - \rho_R$. By studying the long time behaviour of our time evolution equations we find the same SS phases as in \cite{AS}, namely a low--density (LD) phase (with small bulk density $\rho_L$ extending to the left boundary, and a boundary layer, whose characteristic length remains finite in the large $N$ limit, on the right), a high--density (HD) phase (with large bulk density $\rho_R$ extending to the right boundary, and a boundary layer on the left), a maximal current (MC) phase (with bulk density $\rho_* = \argmax J(\rho)$ in the central region of the system and 2 boundary layers) and, for $q$ sufficiently larger than $r$ (numerically we find $q/r \gtrsim 6$) and $1 - \rho_R$ close to $1$, another high--density (labelled HD$'$ in the following) phase. Typical SS phase diagrams are reported in Fig.\ \ref{fig:PhDiagRep} for repulsive interactions ($r > q$), in Fig.\ \ref{fig:PhDiagAttr} for weakly attractive interactions ($q > r$, $q/r$ not too large) and in Fig.\ \ref{fig:PhDiagAttr10} for strongly attractive interactions ($q > r$, $q/r$ large), using solid lines (dashed lines, and the corresponding distinctions between fast and slow phases, will be discussed later). The (continuous) transition line between the LD (respectively HD) and the MC phase is given by the condition $\rho_L = \rho_*$ (resp.\ $\rho_R = \rho_*$), while the (discontinuous) transition line between the LD and HD phases is given by $J(\rho_L) = J(\rho_R)$. The HD$'$ phase appearing in the strongly attractive case in Fig.\ \ref{fig:PhDiagAttr10} has a density profile qualitatively similar to the MC phase, with a central bulk region and 2 boundary layers, but its bulk density $\rho'$ (which depends only on $\rho_R$, as in the ``ordinary'' HD phase) is slightly larger than $\rho_*$ (the largest value found in the case of Fig.\ \ref{fig:PhDiagAttr10} was $0.825 > \rho_* \simeq 0.738$). The (continuous) transition line between the HD$'$ and the MC phases is given by the condition $\rho' = \rho_*$, while the (discontinuous) transition line between the HD$'$ and LD phases is given by $J(\rho') = J(\rho_L)$, but since $\rho'$ is not equal to any of $\rho_*, \rho_L$ and $\rho_R$, we cannot expect its value, and as a consequence the corresponding phase boundaries, to be exact. 

\begin{figure}
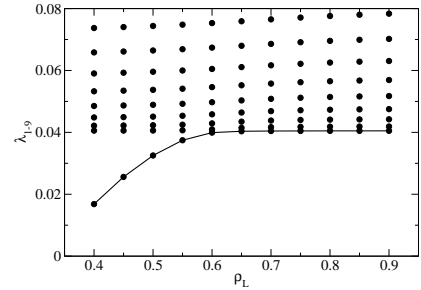

\onefigure[width=0.3\textwidth]{Spectra-N100-q1-r05-rhoR02.eps}
\caption{The bottom part of the spectrum of the relaxation matrix for $N = 100$, $q = 1$, $r = 0.5$ and $\rho_R = 0.8$ (HD phase). Filled circles denote eigenvalues $\lambda_{1-9}$, the line is a guide for the eye connecting relaxation rates $\lambda_1$.}
\label{fig:Spectra}
\end{figure}

We now turn our attention to the investigation of dynamical transitions, which are represented in Figs.\ \ref{fig:PhDiagRep}--\ref{fig:PhDiagAttr10}. Dynamical transitions correspond to singularities (in the infinite size limit) in the relaxation rate $\lambda_1$, without any corresponding singularity in the SS properties. Considering the HD phase to fix ideas, a dynamical transition separates a region of the phase diagram (labelled {\em fast} for reasons which will become clearer in the following) where $\lambda_1$ depends only on $\rho_R$ (the parameter fixing the SS bulk density) from one or more regions (labelled {\em slow}) where $\lambda_1$ depends also on $\rho_L$ (the roles of $\rho_L$ and $\rho_R$ are exchanged in the LD phase). In the case of certain ASEPs, including TASEP (that is the present model with $q = r = 1$) the location of the transition is exactly known, as well as the value of $\lambda_1$ on both sides of the transition \cite{deGierEssler05,deGierEssler08}. The physical meaning of the transition is however not yet clear, as remarked in \cite{ProemeBlytheEvans11} by Proeme, Blythe and Evans. In \cite{TASEP-EPJB} we have shown numerically that the spectrum of the mean--field relaxation matrix at large $N$ has different qualitative properties in the fast and slow phases of TASEP. In the fast phases, as $N \to \infty$, the spectrum tends to a continuous band, while in the slow phases an isolated eigenvalue appears, {\em below} the continuous band, which corresponds to a slowest relaxation mode being {\em much slower} than all the other modes. We have recently confirmed analytically (still at mean--field level) these results \cite{Prep} in the case of both simple TASEP and TASEP with Langmuir kinetics (introduced in \cite{ParmeggianiFranoschFrey03,ParmeggianiFranoschFrey04}) in the so--called balanced case. In the present work we observe the same phenomenon, illustrated in Fig.\ \ref{fig:Spectra}, where we plot the 9 smallest eigenvalues $\lambda_{1-9}$ of the relaxation matrix as a function of $\rho_L$, for $N = 100$, $q = 1$, $r = 0.5$ and $\rho_R = 0.8$, that is in the HD phase in Fig.\ \ref{fig:PhDiagAttr}. One can clearly see a region on the right (the fast phase) where $\lambda_1$ takes its maximum value, independent of $\rho_L$, and a region on the left (the slow phase) where the relaxation is slower and $\lambda_1$ detaches from the rest of the spectrum. One might argue that in the fast phase the left boundary condition is ``consistent'' with the bulk ($\rho_L$ is sufficiently close to the bulk density $\rho_R$), so that the relaxation dynamics is dominated by the bulk properties, while in the slow phase $\rho_L$ is so different from the bulk density $\rho_R$ that the system exhibits a new, boundary--driven, relaxation mode. All the eigenvalues in Fig.\ \ref{fig:Spectra} are real, while going up in the spectrum one encounters also pairs of complex conjugate eigenvalues. 

\begin{figure}
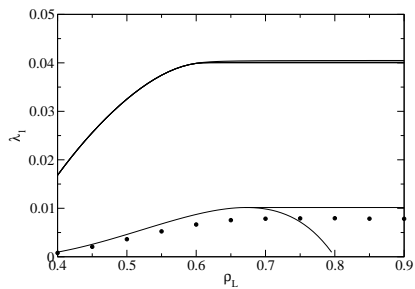

\onefigure[width=0.3\textwidth]{Fig5.eps}
\caption{The relaxation rate $\lambda_1$ as a function of $\rho_L$ for $q = 1$, $r = 0.5$ and $\rho_R = 0.8$ (corresponding to Fig.\ \protect\ref{fig:Spectra}). Thick lines: PA, $N = 100$, 200, 400 and 800 from top to bottom. Thin lines: mDWT. Filled circles: extrapolation of exact finite size results.}
\label{fig:Rate-FSS}
\end{figure}

In Fig.\ \ref{fig:Rate-FSS} the relaxation rate $\lambda_1$ is plotted for various system sizes, for model parameters as in Fig.\ \ref{fig:Spectra}. It is clear that $\lambda_1$ is practically independent of the system size in the slow phase, while some weak size dependence can be observed in the fast phase. This is consistent with the mean--field results in \cite{Prep}, where for the case of pure TASEP the mean--field rate was shown to approach its asymptotic value exponentially (respectively as $1/N^2$) in the slow (resp.\ fast) phase. In the same figure we report, for comparison, results from the mDWT by de Gier and Essler \cite{deGierEssler08}. These authors compared their exact result for the relaxation rate of pure TASEP with the DWT result \cite{Schutz1998,Schutz2000} $\lambda_1 = D_R + D_L - 2 \sqrt{D_L D_R}$, where $D_{L,R} = J(\rho_{L,R})/(\rho_R - \rho_L)$. They found that the DWT result is exact in the slow phase, and the dynamical transition corresponds to a maximum of the DWT rate. In their mDWT, which is exact by construction for pure TASEP, they take the DWT result in the slow phase and the maximum rate in the fast phase. The mDWT is likely to be no longer exact for the AS model, but in Fig.\ \ref{fig:Rate-FSS} we see that it confirms the occurrence of a dynamical transition, whose location is close to the PA one. 
%Moreover, the mDWT results suggest that the PA overestimates rates, as it does in pure TASEP \cite{TASEP-EPJB}. 
Notice also (Fig.\ \ref{fig:PhDiagRep}) that in the repulsive case, as $1 - \rho_R \to 0$, the mDWT predicts that the HD--phase dynamical transition tends to a value $\rho_L < 1$, at odds with the PA. This behaviour is observed for sufficiently strong repulsion, namely $q/r < 0.5$. As a further confirmation, in Fig.\ \ref{fig:Rate-FSS} we plot results obtained along the lines of \cite{Schutz2000,Wehefritz1997} (where very accurate results were obtained for pure TASEP), that is by extrapolating exact finite size ($N \le 24$) results with the Bulirsch--Stoer algorithm \cite{BST1964,Schutz1988}. The parameter $\omega$, characterizing the leading term in the expected size dependence, has been set at 2, based on the exactly known finite size behaviour of the relaxation rate for pure TASEP \cite{deGierEssler05,deGierEssler08}, after verifying numerically that (even for the AS model) this value gives near--optimal results according to the criterion proposed in \cite{Schutz1988}. For $\rho_L \ge 0.7$ the variations in $\lambda_1$ are smaller than $2 \cdot 10^{-4}$, strongly suggesting that the fast phase is not an artifact of the PA and the mDWT.

\begin{figure}
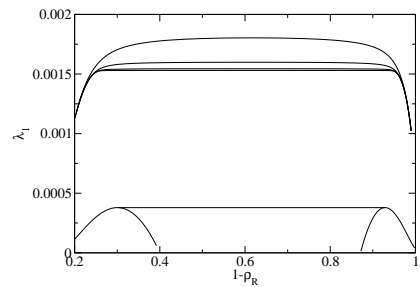

\onefigure[width=0.3\textwidth]{Fig6.eps}
\caption{The relaxation rate $\lambda_1$ as a function of $1-\rho_R$ for $q = 1$, $r = 0.1$ and $\rho_L = 0.6$. Thick lines: PA, $N = 100$, 200, 400 and 800 from top to bottom. Thin lines: mDWT.}
\label{fig:Rate-FSS-Attr10}
\end{figure}

It is a remarkable novel feature of this model that, for strongly attractive interactions as in Fig.\ \ref{fig:PhDiagAttr10}, two dynamical transitions are observed in the LD--phase, with the appearance of 2 LD--slow phases, at small (respectively large) values of $1 - \rho_R$, close to the HD (resp.\ HD$'$) phase. In Fig.\ \ref{fig:PhDiagAttr10} only a portion of the LD--slow phase close to HD$'$ is shown (for the small values of $r$ needed to observe the HD$'$ phase, as $\rho_L$ gets small, the relaxation matrix becomes severely ill--conditioned, and the determination of the dynamical transition line is affected by progressively larger errors). The two dynamical transitions are illustrated in Fig.\ \ref{fig:Rate-FSS-Attr10} by plotting the relaxation rate as a function of $1 - \rho_R$ for various system sizes. The mDWT results confirm the existence of the dynamical transitions also in this case. Here some care is needed for very small $\rho_R$, close to the HD$'$ phase, because the domain wall modelled by the mDWT is between a low--density region of density $\rho_L$ and a high--density, HD$'$--like region whose density is not given by $\rho_R$, but by a function $\rho'(\rho_R)$. Since in PA this function is practically linear, we replaced $\rho_R$ in the mDWT with a linear function fitting the HD$'$ density. For such strongly attractive interactions, the estimates of $\lambda_1$ obtained by extrapolation of exact finite size results are not stable, probably much larger sizes would be needed. 

\begin{figure}
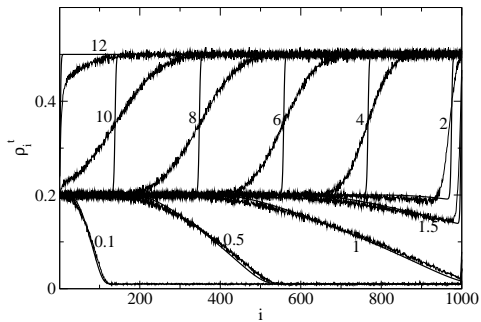

\onefigure[width=0.35\textwidth]{Fig7.eps}
\caption{Density profile as a function of time for $N = 1000$, $q = 0.1$, $r = 1$, $\rho_L = 0.2$, $\rho_R = 0.5$ (HD--slow phase). The number near each line denotes reduced time $t/N$. $t/N = 12$ is indistinguishable from the SS. Thick smooth lines: PA, thin noisy lines: KMC simulation (average over $10^4$ trajectories).}
\label{fig:DynSlow}
\end{figure}

\begin{figure}
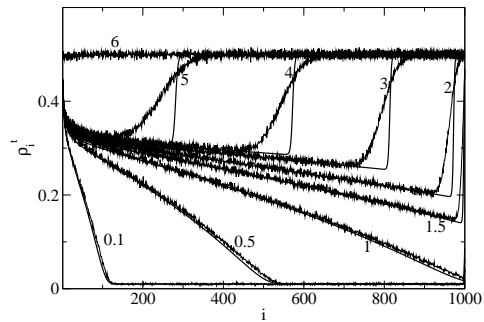

\onefigure[width=0.35\textwidth]{Fig8.eps}
\caption{Same as Fig.\ \protect{\ref{fig:DynSlow}} for $\rho_L = 0.5$ (HD--fast phase). $t/N = 6$ is indistinguishable from the SS.}
\label{fig:DynFast}
\end{figure}

In order to try to understand the physical meaning of the dynamical transition, we have investigated in some detail the full dynamics of the model in the fast and slow phases. In particular, in the repulsive case (Fig.\ \ref{fig:PhDiagRep}), we have analyzed a point in the HD-slow phase ($\rho_L = 0.2$, $\rho_R = 0.5$) and one in the HD-fast phase ($\rho_L = 0.5$, $\rho_R = 0.5$). In both cases we have studied the full time evolution of the density profile, starting from an initial condition with very small (0.01) uniform density and no correlations. Results are reported in Figs.\ \ref{fig:DynSlow} and \ref{fig:DynFast} respectively. In these figures we compare results from the PA with results from kinetic Monte Carlo (KMC) simulations averaging over $10^4$ trajectories, showing that the qualitative picture provided by the PA is correct, the main difference being that shocks are too sharp in the PA (a similar behaviour has been observed in the MCAK results for a slightly more general model \cite{Maass2013}).

In Fig.\ \ref{fig:DynSlow} the dynamics can be divided into 2 parts. In the first part (analogous to the penetration regime in \cite{Maass2013}), until $t_1/N = \rho_L/J(\rho_L) \sim 1.5$, particles fill the lattice (which is initially almost empty) and form an LD--like plateau of density $\rho_L$, which occupies the whole lattice except for a boundary layer near the right end. The second part (analogous to the intermediate regime in \cite{Maass2013}), from $t_1$ to the SS, is characterized by the motion of a shock, separating 2 regions of densities $\rho_L$ and $\rho_R$, respectively. According to the theory of boundary--induced phase transitions \cite{Schutz}, the shock moves leftward with velocity $v_s = (J(\rho_R) - J(\rho_L))/(\rho_R - \rho_L)$.
%, a negative value whose magnitude increases with $\rho_L$. In this particular case we obtain $v_s \simeq -0.105$, and the shock spans the whole lattice in a time $t_2/N = 1/|v_s| \simeq 10$. 

\begin{figure}[h!]
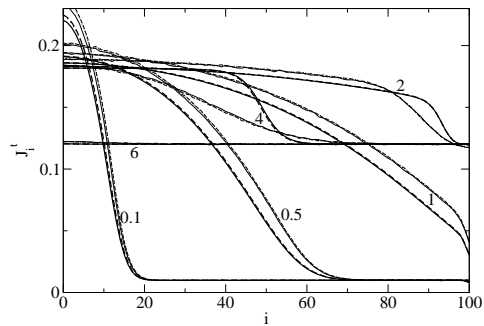

\onefigure[width=0.35\textwidth]{Fig9.eps}
\caption{Current profile as a function of time for $N = 100$, $q = 0.1$, $r = 1$, $\rho_L = 0.5$ (solid lines), $0.7$ (dotted) and $0.9$ (dashed), $\rho_R = 0.5$ (HD--fast phase). The number near each line denotes time $t/N$. $t/N = 6$ is indistinguishable from the SS. Thick smooth lines: PA, thin noisy lines: KMC simulation (average over $10^6$ trajectories).}
\label{fig:DynFastUniv}
\end{figure}

In Fig.\ \ref{fig:DynFast} the dynamics can also be divided into 2 parts. In the first part, however, due to a larger $\rho_L$, the entry rate is so large that the particles do not have time to form a plateau at density $\rho_L$ (this would take a time $t_1/N = \rho_L/J(\rho_L) \sim 4$). When the shock forms it moves certainly faster than in the previous case, since in the slow phase $|v_s|$ was increasing with $\rho_L$. Its speed, which increases with time, is however smaller than $v_s$ (in the limit $\rho_L \to \rho_R^-$), since the density immediately on the left of the shock is smaller than $\rho_L$. Indeed, a more detailed analysis reveals that, in the whole parameter region of the HD--fast phase, the shock speed no longer increases with $\rho_L$. Actually, the full dynamics is practically independent of $\rho_L$, except near the left boundary. This is shown in Fig.\ \ref{fig:DynFastUniv} for the current profiles and, more importantly, this is clearly confirmed by KMC simulations. Similar results are obtained if one considers the density, or the 2--node marginals.

The dynamical features we have obtained above for the HD phases remain valid for other values of $q$ and $r$ and in the LD phases. The HD$'$ phase is characterized by a (very small) relaxation rate which depends on both $\rho_R$ and $\rho_L$, as in the HD--slow phase, but no plateau is formed at intermediate times in the dynamics.

\section{Discussion}

We have considered a simple PA, which extends the 2--node CMF theory and the MCAK by introducing time evolution equations for 2--node expectations, and shares with these techniques the property of reproducing certain exact results for the SS of the AS model with bulk--adapted boundary rates (in particular the bulk current--density relation and the location of most SS phase transitions). We have used this approximation to investigate the relaxation dynamics of the model, finding dynamical transitions similar to those found in ASEPs, both in the LD and HD phases. The existence of these transitions is confirmed by the mDWT by de Gier and Essler and (at least for not too strongly attractive interactions) by extrapolation of exact finite size results. It is remarkable that, for sufficiently strong attractive interactions, two dynamical transitions can be found by PA and mDWT in the same LD phase. 

The dynamical transitions separate slow and fast phases. In the slow phases, the relaxation rate depends on both boundary densities $\rho_L$ and $\rho_R$, while in the fast phases it depends only on the parameter which determines the bulk density, that is $\rho_L$ in the LD phase and $\rho_R$ in the HD phase. We have shown, confirming results we had already obtained \cite{TASEP-EPJB, Prep} in TASEP with various mean--field like approximations, including the PA, that the spectrum of the relaxation matrix changes qualitatively at a dynamical transition. In the fast phase, it tends to a continuous band, while in the slow phase, an isolated eigenvalue, corresponding to the relaxation rate appears below the continuous band. A natural interpretation is that in the fast phase the boundary condition which does not determine the bulk density (e.g.\ $\rho_L$ in the HD phase) is ``consistent'' with the bulk ($\rho_L$ is sufficiently close to the bulk density $\rho_R$), so that the relaxation dynamics is dominated by the bulk properties (hence becoming independent of $\rho_L$), while in the slow phases one boundary density is so different from the bulk density that a slower, boundary--driven, relaxation mode appears. 

We have also studied the full relaxation dynamics in the slow and fast phases, looking for qualitative differences. An interesting result is that in the HD--slow (respectively LD--slow) phases, with initial conditions corresponding to an almost empty (resp.\ almost full) lattice, the system develops an LD--like (resp.\ HD--like) plateau before reaching the SS. This plateau is not a long--lived metastable state, nevertheless, since the slow phases are located near the LD--HD transition lines, on which these phases coexist, it is tempting to view the slow phases as (loose) analogues of metastability regions in an equilibrium phase diagram. No such plateaus are observed in the fast phases, and another remarkable result is that in these phases the full dynamics, not just the relaxation rate, depends only on the parameter which determines the bulk density, as shown in Fig.\ \ref{fig:DynFastUniv} considering current profiles. In the same figure we have also reported Kinetic Monte Carlo simulation results, which confirm that the full dynamics is independent of $\rho_L$. 
A direct calculation of the relaxation rate with KMC would also be welcome, in order to confirm the results illustrated in Figs.\ \ref{fig:Rate-FSS} and \ref{fig:Rate-FSS-Attr10}, but unfortunately this seems not feasible, as discussed in \cite{ProemeBlytheEvans11}, where the authors eventually switched to a density--matrix renormalization group approach. After simulating systems of sizes up to $N = 1000$ (much larger than in \cite{ProemeBlytheEvans11}), we similarly observe that it is very difficult to get a clear single--exponential relaxation. One can reasonably argue that this is to be ascribed to the small separation between the lowest eigenvalues of the relaxation matrix. 
We believe, however, that the PA and mDWT results for the relaxation rate, supported by extrapolation of exact finite size results (which for pure TASEP is at least as accurate as the density--matrix renormalization group) make a strong case in favor of the existence of dynamical transitions in the AS model. Furthermore, the agreement of the PA with the KMC results, confirming in particular that the full dynamics is independent of $\rho_L$ in the HD--fast phase, support the overall reliability of the PA results. 

Work is in progress to extend these results to other models, and we hope that the present paper will stimulate further investigations about the possible onset of dynamical transitions in non--equilibrium SS, including those exhibited by more realistic traffic models like the ones considered in \cite{Maass2011,Maass2012,Maass2013,Kolomeisky2015A,Kolomeisky2015B,Baumgaertner2017,Kolomeisky2018A,Kolomeisky2018B}.

\end{document}